%% file: LongAsymmPbPb-2.76TeV.tex
\documentclass[ALICE,manyauthors]{cernphprep}

\usepackage{lineno}
\usepackage{amsmath}
\usepackage[comma,square,numbers,sort&compress]{natbib}
\usepackage{hyperref}
\usepackage{graphicx}  
\usepackage{dcolumn}   
\usepackage{bm}        
\usepackage{amssymb}   
\usepackage{amsfonts}
\usepackage{graphics}
\usepackage{grffile}   
\usepackage{epsfig}
\usepackage{units}
\usepackage{comment}
\usepackage[usenames]{color}
\usepackage[normalem]{ulem} 
\usepackage{color}
\usepackage[utf8]{inputenc}
\usepackage[T1]{fontenc}
\usepackage{caption}
\usepackage{authblk}

\graphicspath{{./pics/}}

\newcommand{\com}[1]       {\relax}
\newcommand{\PbPb}         {\mbox{Pb--Pb}{ }}

\newcommand{\pPb}          {\mbox{p--Pb}}

\newcommand{\dAu}          {\mbox{d--Au}}

\newcommand{\dNdeta}    {\mathrm{d}N/\mathrm{d}\eta}

\newcommand{\snn}          {\ensuremath{\sqrt{s_{\mathrm{NN}}}}}
\newcommand{\snnbf}        {\ensuremath{\mathbf{{\sqrt{\textit{s}_{NN}}}}}}

\newcommand{\blue}[1]      {\textcolor{blue}{#1}}

\begin{document}
\begin{titlepage}
\PHyear{2017}
\PHnumber{277}      
\PHdate{16 October}  

\title{Longitudinal asymmetry and its effect on pseudorapidity
  distributions \\
in \PbPb collisions at \snnbf  
 = 2.76 TeV}
\ShortTitle{Longitudinal asymmetry and $\dNdeta$ 
in \PbPb collisions at \snnbf ~= 2.76 TeV}   

\Collaboration{ALICE Collaboration\thanks{See Appendix~\ref{app:collab} for the list of collaboration members}}
\ShortAuthor{ALICE Collaboration} 


\begin{abstract}

First results on the longitudinal asymmetry and its effect on the
pseudorapidity distributions in \PbPb collisions at \snn = 2.76 TeV at
the Large Hadron Collider are obtained with the ALICE detector.
The longitudinal asymmetry arises because of an unequal
  number of participating nucleons from the two colliding nuclei, and is 
  estimated for each event by measuring the energy in the forward
  neutron-Zero-Degree-Calorimeters (ZNs). The effect
  of the longitudinal asymmetry is measured on the pseudorapidity distributions of
  charged particles in the regions $|\eta| < 0.9$,  $2.8 < \eta < 5.1$ and
  $-3.7 < \eta < -1.7 $ by taking the ratio of the pseudorapidity distributions from events
corresponding to different regions of asymmetry.  The coefficients of
a polynomial fit to the ratio characterise the effect of the
asymmetry. A Monte Carlo simulation using a Glauber
  model for the colliding nuclei is
  tuned to reproduce the spectrum in the ZNs and provides 
  a relation between the measurable longitudinal asymmetry and the shift in
  the rapidity ($y_{\mathrm{0}}$) of the participant zone formed by the
  unequal number of participating nucleons.
The dependence of the coefficient of the linear term in the polynomial
expansion, $c_{\rm 1}$, on the mean value of $y_{\mathrm{0}}$ is investigated.

\end{abstract}
\end{titlepage}

\setcounter{page}{2}
\section{Introduction}
\label{sec:intro}
In a heavy-ion collision, the number of nucleons participating from
each of the two colliding nuclei is finite, and will
fluctuate event-by-event. The kinematic centre of mass of the
participant zone, defined as the overlap region of the colliding
nuclei, in general  
has a finite momentum in the nucleon-nucleon centre of 
mass  frame because of the unequal number of nucleons participating
from the two nuclei. This
momentum causes a longitudinal asymmetry in the collision and
corresponds to a shift of rapidity of the participant zone with respect 
to the nucleon-nucleon centre of mass (CM) rapidity, termed the
rapidity-shift $y_{\mathrm{0}}$. The value of $y_{\rm 0}$ 
is indicative of the magnitude of the longitudinal asymmetry of the
collision ~\cite{Vovchenko:2013viu, Raniwala:2016ugm}. Assuming the number
of nucleons participating from each of the two nuclei is A and B, 
the longitudinal asymmetry in participants is defined as $\alpha_{\rm {part}} = \frac{A-B}{A+B}$
and the rapidity-shift can be approximated as $y_{\rm 0} \cong
\frac{1}{2}ln\frac{A}{B}$ at LHC energies~\cite{Raniwala:2016ugm}.

The shift in
the CM frame of the participant zone, which evolves into a state of
dense nuclear matter, needs to be explored in heavy-ion collision models. 
Comparison of model predictions with the observed $\Lambda$-polarisation, possibly due to vorticity
from the initial state angular momentum surviving the evolution,
requires a precise determination of initial conditions and hence the
shift in the CM frame~\cite{STAR:2017ckg, Becattini:2013vja, Xie:2017upb}. 
Such a shift may also affect
observations on correlations amongst particles, which eventually
provide information about the state of the matter through model
comparisons. Further, the resultant decrease in 
the CM energy may affect various observables including the particle
multiplicity. 
The transverse spectra are known to be affected by the initial geometry of the
events, as estimated through techniques of event shape engineering,
indicating an interplay between radial and transverse flow~\cite{Adam:2015eta}. 
The
measurement of 
longitudinal asymmetry will provide a new parameter towards event
shape engineering, affecting many other observables. 

The simplest of all possible investigations into the effect of longitudinal asymmetry
is a search for modification of the kinematic distribution of the particles. The pseudorapidity distribution ($\dNdeta$) of soft particles,
averaged over a large number of events, is 
symmetric in collisions of identical nuclei. 
These distributions were observed to be asymmetric in collisions of unequal nuclei such as  \dAu~\cite{Back:2003hx} and 
\pPb~\cite{ALICE:2012xs, Adam:2014qja, Aad:2015zza} and have been explained  in terms of the rapidity-shift of the participant zone~\cite{Steinberg:2007fg}.
In a heavy-ion collision, the effect of the rapidity-shift of the participant zone should be
discernible in the distribution of produced particles. This small effect can be
estimated by taking the ratio of pseudorapidity distributions in
events corresponding to different longitudinal asymmetries~\cite{Raniwala:2016ugm}.

It was suggested that the rapidity
distribution of an event, scaled by the average rapidity distribution,
can  be expanded in terms of Chebyshev 
polynomials, where the coefficients of expansion are measures of the
strength of longitudinal fluctuations and can be determined by
measuring the two particle correlation function~\cite{Bzdak:2012tp}. Using the same methodology,
the event-by-event pseudorapidity distributions are also expanded 
in terms of Legendre polynomials~\cite{Jia:2015jga}.
The ATLAS collaboration expanded the pseudorapidity 
distributions in terms of Legendre polynomials and obtained the
coefficients by studying pseudorapidity
correlations~\cite{Aaboud:2016jnr}.

In the present work, the events are classified according to 
the asymmetry determined from the measurement of energies of neutron
spectators on both sides
of the collision  ~\cite{Raniwala:2016ugm}.
The effect of asymmetry
is investigated by taking the ratio of the measured raw $\dNdeta$ distributions 
for events from different regions of the distribution of
measured asymmetry. A major advantage of studying this ratio
is the cancellation of (i) systematic uncertainties and (ii) the effects of short range correlations.
The first measurements of the effect of
asymmetry on the raw $\dNdeta$ distributions are reported here.

The paper is organised as follows: Sect.~\ref{sec:expdetail} provides an introduction to the 
experimental setup and the details of the data sample.
Section~\ref{sec:Analysis} discusses the characterisation of the change in raw $\dNdeta$
distributions for events classified in 
different asymmetry regions. Section~\ref{sec:Simula} describes the simulations
employed to provide a relation between 
the measured asymmetry and the rapidity-shift $y_{\rm 0}$
of the participant zone. 
The relation between the parameter characterising the change
in raw $\dNdeta$ distributions is shown for different centralities in
Sect.~\ref{sec:Result}, along with its relation to the estimated values of  $y_{\mathrm{0}}$.

\section{Experimental details and data sample}
\label{sec:expdetail}

The analysis uses data from \PbPb collision events at \snn
=  2.76 TeV, recorded in the ALICE experiment in 2010, with a minimum
bias trigger ~\cite{Aamodt:2010pa, Aamodt:2010cz}. The data used in
the present analysis is recorded in the neutron Zero Degree
Calorimeters (ZNs), the V0 detectors, the Time
Projection Chamber (TPC) and the Inner Tracking System (ITS). Both ZNs
and V0 detectors are on either side of the interaction vertex, those
in the direction of positive pseudorapidity axis are referred as
V0A and ZNA and those in the opposite direction are referred as V0C
and ZNC. A detailed 
description of the ALICE detectors and their performance can be found elsewhere
~\cite{Aamodt:2008zz, Abelev:2014ffa}.

The event asymmetry is estimated using the energy measured in the two
ZNs situated 114 metres away from the nominal interaction point (IP) on
either side.
The ZNs detect only spectator neutrons that are not bound in nuclear
fragments, since the latter are bent away by the magnetic field of the LHC separation dipole. The ZN
detection probability for neutrons is 97.0\% $\pm$ 0.2\%(stat)
$\pm$3\%(syst) ~\cite{ALICE:2012aa}. The relative energy resolution of
the 1{\it n} peak at 1.38 TeV is 21\% for the ZNA and 20\%
for the ZNC ~\cite{ALICE:2012aa}.
The production of nuclear fragments increases with collision impact parameter
degrading the resolution on the number of participating nucleons.
The energy in the ZNs is a good measure of the number of
spectator neutrons only for the more central
collisions~\cite{Abelev:2014ffa}. The analysis is limited to the top 35\% most central
sample and employs data from $\sim 2.7$ million events.

The raw $\dNdeta$ distributions in the region $|\eta| < 0.9$ are obtained by
reconstructing the charged particle tracks using the TPC and ITS.
The
requirements on the reconstructed tracks obtained using the
measurements in these
detectors are the same as
in other earlier analyses \cite{Aamodt:2010pa}.
The measured amplitudes in the 
V0A ( $+2.8 < \eta < +5.1$) and V0C ($-3.7 < \eta < -1.7$ ) are used
to estimate the raw $\dNdeta$ distributions of charged particles in the forward
regions. Both V0A and V0C are scintillator
counters, each with four segments in pseudorapidity and eight segments
in azimuth.  The raw distributions measured are termed as $\dNdeta$ distributions
throughout the manuscript. 
In order to ensure a uniform detector performance,  
the present analysis uses events with z position (along the beam
direction) of the interaction vertex, $V_{\rm z}$, within
$\pm$ 5 cm of the IP in ALICE.  
The centrality of Pb-Pb collisions was estimated by two independent
methods. One estimate was based on the charged particle multiplicity
reconstructed in the TPC and the other was based on the 
amplitudes in the V0 detectors \cite{Abelev:2013qoq}.

\section{Analysis and systematic uncertainties}
\label{sec:Analysis}

In the present analysis, changes in the raw pseudorapidity distribution of
charged particles are investigated for different values of measured
asymmetry of the event. 
The method of measurement of the asymmetry and the parameters
characterising the change in $\dNdeta$ distributions are discussed in
this section.

\subsection{Analysis} 

Any event asymmetry due to unequal number of nucleons
from the two participating nuclei may manifest itself in the longitudinal distributions,
i.e. $\mathrm{d}N/\mathrm{d}y$ (or $\dNdeta$) of
the produced particles because of a shift in the effective CM.
Assuming that the rapidity distributions can be described by a
symmetric function about a mean $y_{\rm 0}$ 
($y_{\rm 0}$ = 0.0 for symmetric events), the ratio of the distributions for
asymmetric and symmetric events may be written as
\begin{equation}
\begin{split}
\frac {(dN/dy)_{\rm {asym}}}{(dN/dy)_{\rm {sym}}}  = \frac {f
  (y-y_{\rm 0})} { f(y)}
\propto \sum_{n=0}^{\infty} c_{\rm n}(y_{\rm 0})y^{\rm n} \\
\end{split}
\label{eq:RatioEqn}
\end{equation}
For any functional form of the rapidity distribution, this ratio may
be expanded in a Taylor series. The coefficients 
$c_{\rm n}$ 
of the different terms in the expansion depend on the shape and the
parameters of the rapidity distribution ~\cite{Raniwala:2016ugm}. In the ALICE experiment, the
pseudorapidities of the emitted particles were measured. The effect of a rapidity-shift $y_{\rm 0}$ on
the pseudorapidity distribution is discussed in
Sect.~\ref{rapshift}.

The unequal number of participating nucleons will yield a non-zero $y_{\rm 0}$
of the participant zone and will cause an asymmetry in the number of
spectators. This asymmetry can provide information about the
mean values of $y_{\rm 0}$ using the response matrix discussed in Sect.~\ref{sec:Simula}. 
The asymmetry of each event is estimated by
measuring the energy in the ZNs on both sides of the
interaction vertex: $E_{\rm {ZNA}}$ on the side referred to as the A-side
($\eta > 0 $) and $E_{\rm {ZNC}}$ on the side referred to as the 
C-side ($\eta < 0 $). A small difference in the mean and the
relative energy resolution of the 1{\it n} peak at 1.38 TeV  was observed in the performance of
the two ZNs ~\cite{ALICE:2012aa}. For each centrality interval, the
energy distribution in each ZN is divided by its mean, and
the width of the $E_{\rm {ZNC}}/\langle E_{\rm {ZNC}} \rangle$
distribution is scaled to the width of the corresponding distribution
using $E_{\rm {ZNA}}$. The asymmetry in ZN is defined as 
\begin{equation}
\alpha_{\rm {ZN}} = \frac{\epsilon_{\rm {ZNA}}-\epsilon_{\rm
    {ZNC}}}{\epsilon_{\rm {ZNA}}+\epsilon_{\rm {ZNC}}} 
\end{equation} 
where $\epsilon_{\rm {ZNC(A)}}$ is a dimensionless quantity for each event, obtained after scaling the
distributions of $E_{\rm {ZNC(A)}}$ as described above.

\begin{figure}[ht!]
\begin{center}
\includegraphics[trim= 0.0cm 0.55cm 0.0cm 1.05cm,clip, width=0.6\textwidth]{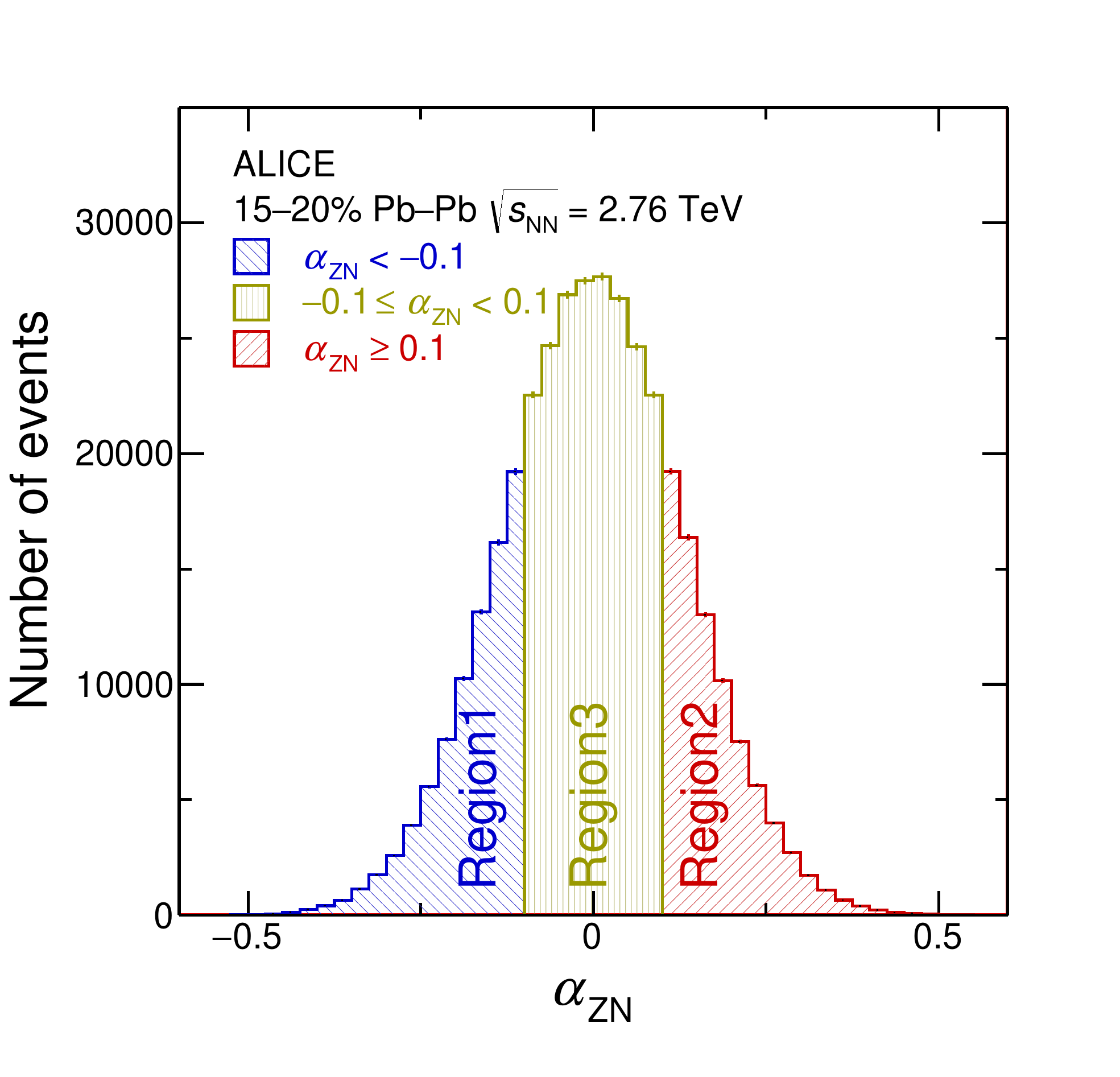}
\caption{The distribution of the asymmetry parameter $\alpha_{\rm{ZN}}$ for the 15--20\% centrality interval. The distribution is demarcated into
  three regions using $|\alpha_{\rm {ZN}}^{\rm {cut}}|$. A Gaussian
  fit to the distribution yields a width of 0.13.}
\label{fig:AlphaDisbnData}
\end{center}
\end{figure}

For the 15--20\% centrality interval, Fig.~\ref{fig:AlphaDisbnData} shows the distribution of the asymmetry $\alpha_{\rm{ZN}}$.
\begin{figure}[ht!]
\begin{center}
\includegraphics[trim= 0.0cm 0.2cm 0.0cm 0.0cm,clip,width=0.9\textwidth]{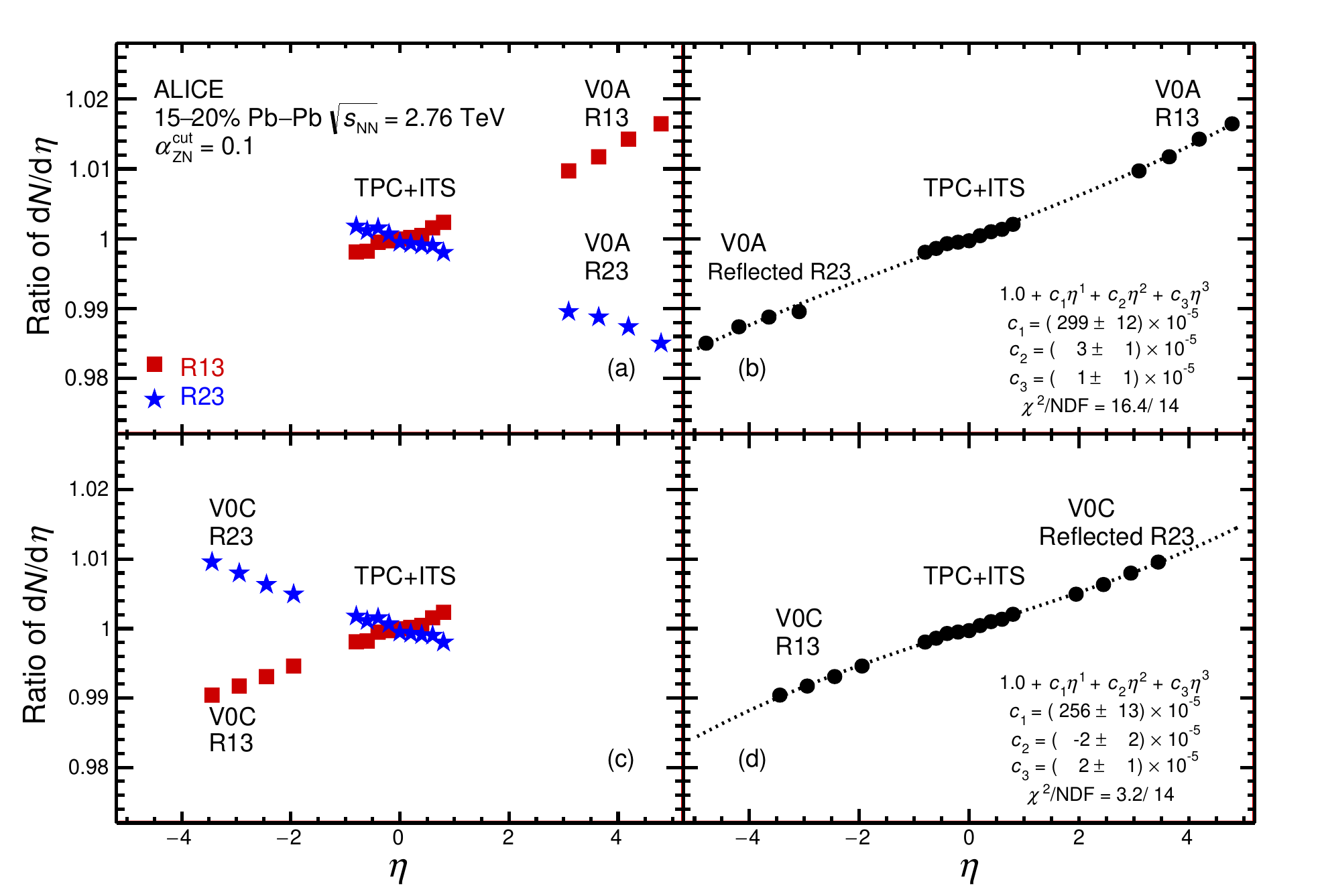}
\caption{
The ratio of $\dNdeta$ distribution for events from the different regions of 
$\alpha_{\rm{ZN}}$ distribution of Fig.~\ref{fig:AlphaDisbnData}. The  $\dNdeta$ distributions are
obtained as described in Sect.~\ref{sec:expdetail}. (a) The square
  (star) symbols corresponding
  to R13 (R23) are obtained by taking the ratio of $\dNdeta$ of
 events from Region 1 (Region 2) to Region 3. (b) The data points are
 obtained after reflection across $\eta = 0$ as described in the text.
The data for $ |\eta| > 1.0 $ in panels (a) and (b) are from
measurements in V0A and in panels (c) and (d) are from measurements in
V0C.}
\label{fig:RatiodNdetaV0A}
\end{center}
\end{figure}
To investigate the significance of this distribution, the
contribution of the resolution of ZNs to the resolution of the
asymmetry parameter $\alpha_{\rm {ZN}}$ is evaluated. For each centrality interval, values of
$E_{\rm {ZNC}}$ and $E_{\rm {ZNA}}$ are simulated for each event by assuming a
normal distribution peaked at the mean value corresponding to the average number of
neutrons and the corresponding energy resolution. The average number
of neutrons is estimated by dividing the experimental distribution of
energy in ZN by 1.38 TeV. These values are
used to obtain $\alpha_{\rm{ZN}}$ for each event and its distribution.  The width
of the distribution corresponds to the intrinsic resolution of the measured
parameter $\alpha_{\rm{ZN}}$ and
varies from 0.023 to
0.050 from the most peripheral (30--35\%) selection to the most
central (0--5\%) selection. The observed width of 0.13 of the
distribution of $\alpha_{\rm {ZN}}$ reported in Fig.~\ref{fig:AlphaDisbnData} is
considerably larger than the resolution of $\alpha_{\rm {ZN}}$ (0.027
for the centrality interval corresponding to the data in the figure)
and the increase in width may be attributed to  
the event-by-event fluctuations in the number of neutrons 
detected in each ZN. To investigate the effect of $\alpha_{\rm{ZN}}$ on
the $\dNdeta$ distributions, the events are demarcated into three
regions of asymmetry by choosing a cut value $\alpha_{\rm{ZN}}^{\rm{cut}}$.
These regions correspond to (i)  $\alpha_{\rm {ZN}}
< -\alpha_{\rm {ZN}}^{\rm {cut}} $ (Region 1), (ii) $\alpha_{\rm {ZN}}
\geq \alpha_{\rm {ZN}}^{\rm {cut}} $ (Region 2)  and
 (iii) $-\alpha_{\rm {ZN}}^{\rm {cut}}\leq \alpha_{\rm {ZN}} < \alpha_{\rm {ZN}}^{\rm {cut}}$
(Region 3). 
Regions 1 and 2 are
referred to as the asymmetric regions and Region 3 is referred to as the
symmetric region.

The effect of the measured asymmetry $\alpha_{\rm{ZN}}$ on the pseudorapidity
distributions is investigated by studying
the ratio of  $\dNdeta$ distribution in events from the asymmetric
region to those from the symmetric region.  There are small differences in the distributions of centrality
and vertex position in events of different regions of asymmetry. 
It is necessary to ensure that any correlation between the
ratio of  $\dNdeta$ and the asymmetry is not due to a systematic effect of
a shift in the interaction vertex. To eliminate any possible systematic bias on the measured
distributions,  the  $\dNdeta$ distributions are corrected by weight factors obtained
by normalising the number of events
in asymmetric and symmetric regions in each 1\% centrality
interval and each 1 cm range of vertex positions.

For the 15-20\% centrality interval, the distributions of
these factors in the two cases corresponding to the asymmetry
regions 1 and 2 have a mean of 1.0 and an rms of 0.05 and 0.06
respectively. The weight factors do not show any systematic dependence
on the position of the vertex. This is expected
considering the large distance between the ZNs as compared to
variations in the vertex position. 
The factors show a systematic
dependence on 1\% centrality bins within each centrality interval.
The 1\% centrality bin with the greater
number of participants tends to have more asymmetric events, presumably to
compensate for the decrease in the effective CM energy due to the
motion of the participant zone; the weight factor is 1.08
for the most central 15--16\% centrality bin and is 0.94 for the
19--20\% centrality bin.


The ratio of $\dNdeta$ for events
corresponding to different regions of asymmetry, as shown in Fig.~\ref{fig:AlphaDisbnData}, is determined.  For $|\eta| < 1.0$, the ratio is obtained
using $\dNdeta$ for tracks. For $|\eta| > 1.0$, the ratio shown in
Fig.~\ref{fig:RatiodNdetaV0A}(a) and (b) is
obtained from amplitudes measured in V0A and the one shown in  
 Fig.~\ref{fig:RatiodNdetaV0A}(c) and (d) is from amplitudes measured
in V0C. The squares in 
Fig.~\ref{fig:RatiodNdetaV0A} (a) and (c) represent the ratio of $\dNdeta$
in the asymmetry Region 1 to that in Region 3 (R13), and the stars represent
the corresponding ratio in Region 2 to Region 3 (R23).
 The filled circles in Fig.~\ref{fig:RatiodNdetaV0A} (b) and (d) are obtained by (i) reflecting the  data points
 labelled R23 across $\eta = 0$ and (ii) taking the averages of R13 and
 reflected-R23 for $|\eta| < 1.0 $.
A third order polynomial is fitted to the points and the
values of the coefficients  $c_n$ along with the $\chi^{\rm 2}$ are 
shown. The polynomial fit to the ratio of  $\dNdeta$ distribution has a dominantly linear term.
A small residual detector effect is observed when determining $c_{1}$
using data measured in V0A and when using data measured in V0C. 
In all subsequent discussion, the values of $c_{\rm 1}$ quoted are the mean of
values obtained from the measurements in V0A and V0C.

\begin{figure}[ht!]
\begin{center}
\includegraphics[width=0.6\textwidth]{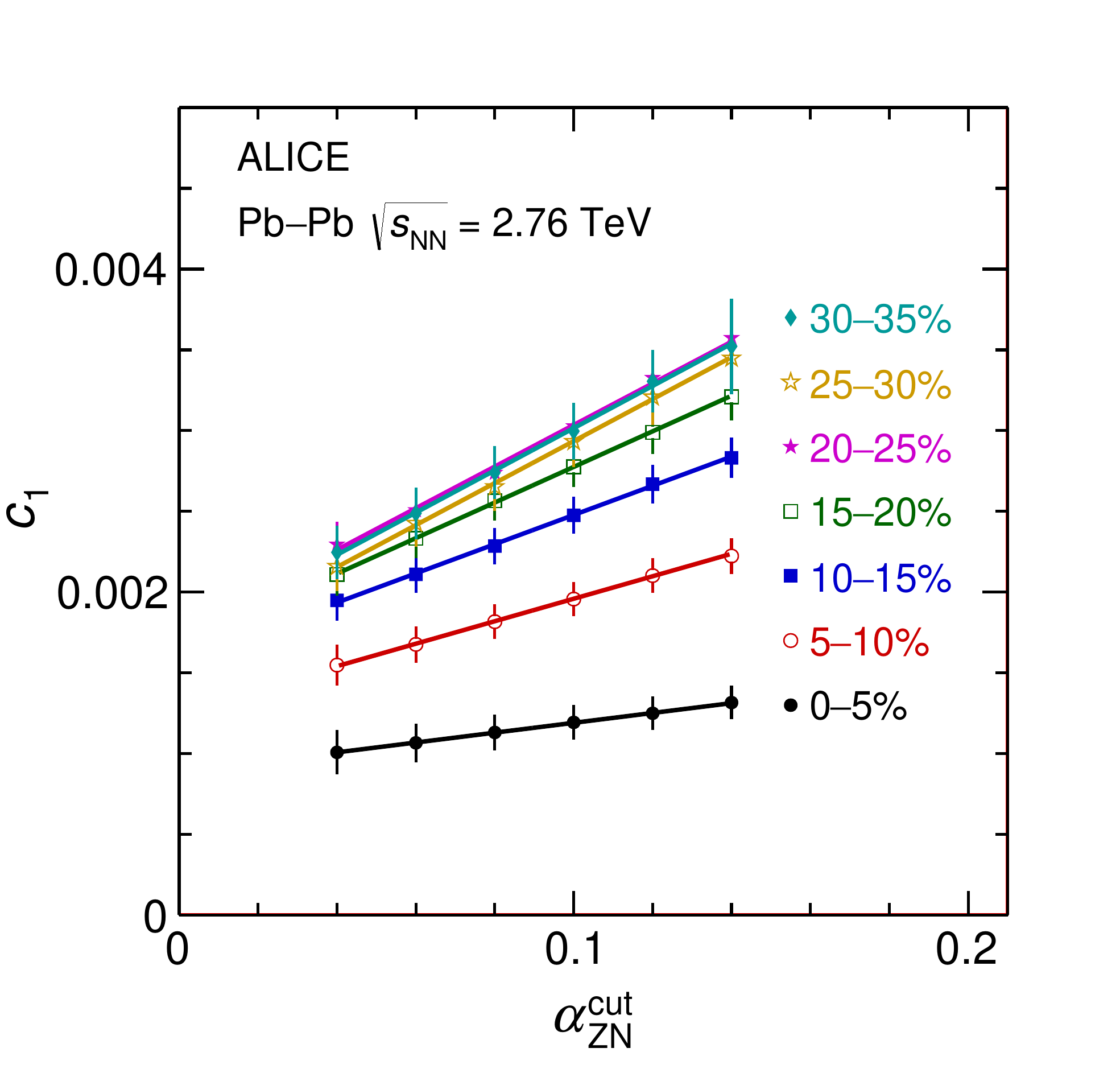}
\caption{ 
  The coefficient $c_{\rm 1}$ characterising the change in $\dNdeta$
  distribution for asymmetric regions is shown for different values of
  $\alpha_{\rm{ZN}}^{\rm{cut}}$ ($\alpha_{\rm{ZN}}^{\rm{cut}}$ demarcates the asymmetric and symmetric events)
  for each centrality interval.}
\label{fig:c1alphazdc}
\end{center}
\end{figure}

Considering that the event samples corresponding to different
regions of asymmetry are identical in all aspects other than their
values of measured $\alpha_{\rm{ZN}}$, the observation of non-zero values of
$c_{\rm 1}$ can be attributed to the asymmetry. 
For a fixed
centrality interval, $c_{\rm 1}$  depends on the choice of
$\alpha_{\rm{ZN}}^{\rm{cut}}$. The analysis is repeated for different values
of $\alpha_{\rm{ZN}}^{\rm{cut}}$ and the dependence of $c_{\rm 1}$ on
$\alpha_{\rm{ZN}}^{\rm{cut}}$ is shown in Fig.~\ref{fig:c1alphazdc}, for
different centralities. 
For each centrality interval the
coefficient $c_{\rm 1}$ has a linear dependence on
$\alpha_{\rm{ZN}}^{\rm{cut}}$ and the slope increases with decreasing
centrality; $c_{\rm 1}$
increases for events corresponding to larger values of average event asymmetry. 
The range of values of $\alpha_{\rm{ZN}}^{\rm{cut}}$
was guided by the resolution and the width of the distribution of
$\alpha_{\rm{ZN}}$, as mentioned in reference to Fig.~\ref{fig:AlphaDisbnData}. 
Increasing the value of  $\alpha_{\rm{ZN}}^{\rm{cut}}$ increases the
mean $\langle \alpha_{\rm {ZN}} \rangle$ for events from
the asymmetric class (Region 1 or Region 2), and increases the RMS of
$\alpha_{\rm{ZN}}$ for events from the symmetric class
(Region 3).

\subsection{Systematic uncertainties} 

The current method of analysis uses the ratio of
two $\dNdeta$ distributions from events divided on the basis of
measurements in ZNs, within a centrality interval. All effects
due to limited efficiency, acceptance or contamination would cancel while obtaining the value of the ratio.
The
contributions to the systematic uncertainties on  $c_{\rm 1}$  are estimated due to the following sources: 

\begin{enumerate}

\item Centrality selection: the ratio of $\dNdeta$ is obtained
  from the measurements of tracks in the ITS+TPC at midrapidity and charge
  particle signal
  amplitudes in the V0 at
  forward rapidities. For the former, the event centrality is
  determined using the measurements in the
  V0 and for the latter using the track multiplicity in the TPC.  The analysis is repeated by 
  interchanging the centrality criteria and
  the resultant change in the values of $c_{\rm 1}$ for different
  centrality intervals is in the range 0.1\% to 3.6\%.  

\item V0A and V0C: the systematic uncertainty on the mean value of $c_{\rm 1}$ is estimated by
  assuming a uniform probability distribution for the correct value of
  $c_{\rm 1}$ to lie between the 
 two values obtained using the charged-particle signal amplitudes
 measured in the V0A and the V0C. The uncertainty is in 
  the range 2.1\% to 4.6\% and does not depend on the centrality
  value.

\item Vertex position: the analysis is repeated  for the z 
  position of the interaction vertex $|V_{\rm z}| \leq
  3.0 $ cm. For the most central interval, the results change by less
  than 0.1\%.  For the 15--20\% centrality interval, the results
  change by 3.3\% and for all
  other centrality intervals, the changes are less than 1.3\%.  

  \item  Weight factors for normalisation: the analysis is also
    repeated without the weight factors mentioned in Sect. 3.1
    for the centrality and the vertex normalisation in the number of
    events. The change in the results is 4.9\% in the most central class and less
    than 1\% for all other centrality intervals.

\end{enumerate}

The total systematic uncertainty is obtained by adding the four uncertainties in
quadrature. The resultant uncertainty is in the range 2.3\% to 5.8\% and is
shown by the band in Fig.~\ref{fig:Meanc1centrality}.


\section{Simulations}
\label{sec:Simula}

The simulation used for obtaining a relation between 
rapidity-shift $y_{\rm 0}$ and the measurable asymmetry $\alpha_{\rm{ZN}}$ is described in
this section.
This simulation has three components: (i) a Glauber Monte Carlo to
generate number of participants and spectator protons and
neutrons, (ii) a
function parametrised to fit the average loss of spectator neutrons
due to spectator fragmentation (the loss of spectator neutrons in each event is smeared around this
average) and (iii) the
response of the ZN to single neutrons. The simulation encompassing
the above is referred to in the
present work as Tuned Glauber Monte Carlo (TGMC), and reproduces the
energy distributions in the ZNs. 
The effect of $y_{\rm 0}$ on the pseudorapidity
distributions has been estimated using additional simulations for a
Gaussian ${\mathrm{d}N/\mathrm{d}y}$ and are also described in this section.

\subsection{Asymmetry and rapidity-shift}
\label{subsec:ZNasymtoy0}

The Glauber Monte Carlo model~\cite{Alver:2008aq} used in the present work assumes a
nucleon-nucleon interaction cross section of 64 mb at \snn~~=
2.76 TeV. The model yields the number of participating nucleons in
the overlap zone from each of the colliding nuclei. 
The range of impact parameters for each 5\% centrality
interval is taken from our Pb-Pb centrality paper~\cite{Abelev:2013qoq}. 
For each centrality interval, 0.4 million events are generated.

\begin{figure}[ht!]
\begin{center}
\includegraphics[width=1.0\textwidth]{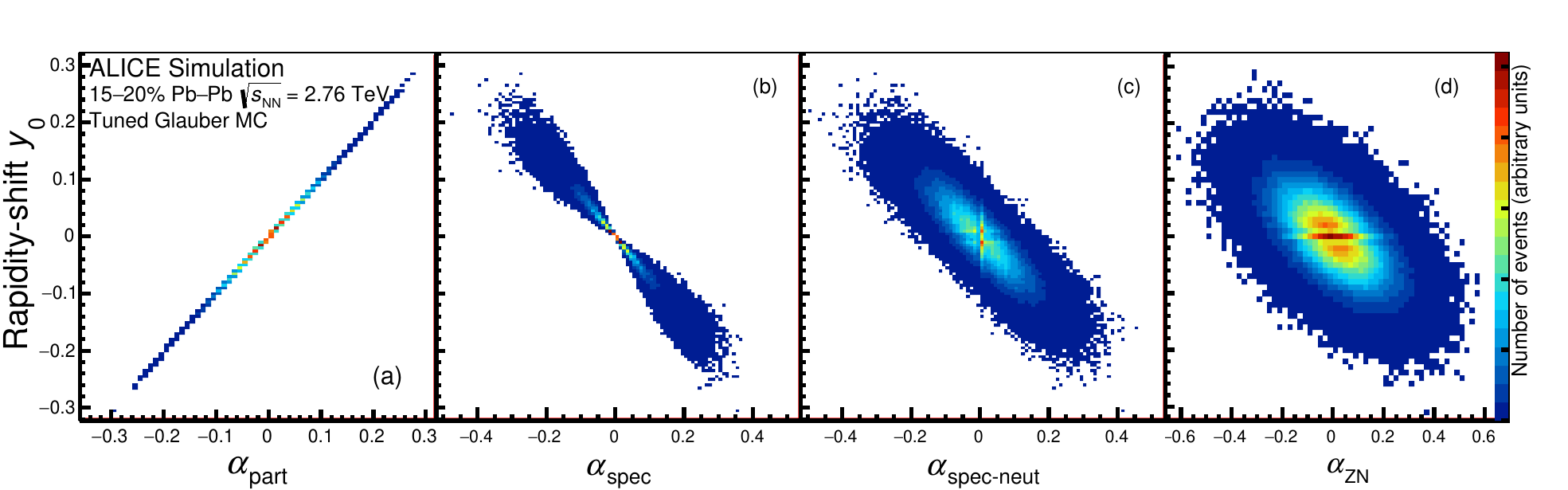}
\caption{Rapidity-shift $y_{\rm 0}$ as a function of asymmetry in (a) number
  of participants (b) number of spectators (c) number of
  spectator neutrons and (d) energy in
  ZN obtained using TGMC as described in the text. The
results in all four panels are shown for the 15--20\% centrality interval.}
\label{fig:GlauberMCy0}
\end{center}
\end{figure} 
For each generated event, the number of participating protons and
neutrons is obtained, enabling a determination
of the
rapidity-shift $y_{\rm 0}$ and the various longitudinal asymmetries. If A and B are the number of
spectators (spectator neutrons) in the two colliding nuclei, the
asymmetry is referred to as $\alpha_{\rm spec}$
($\alpha_{\rm {spec-neut}} $). 
Figure~\ref{fig:GlauberMCy0} (a)
shows the correspondence between $y_{\rm 0}$ and
$\alpha_{\rm {part}}$. Figures~\ref{fig:GlauberMCy0} (b) and (c) show the relation between $y_{\rm 0}$
and $\alpha_{\rm {spec}}$ and $\alpha_{\rm {spec-neut}}$ respectively\blue{~\cite{Raniwala:2016ugm}}. 
These figures show that the rapidity-shift $y_{\rm 0}$ can be estimated by
measuring $\alpha_{\rm {spec}}$ or
$\alpha_{\rm {spec-neut}}$ in any experiment that uses Zero Degree Calorimeters. However, the lack of information
on the number of participants worsens the precision in determining
$y_{\mathrm{0}}$. Figure~\ref{fig:GlauberMCy0} (d) shows the relation between $y_{\rm 0}$
and $\alpha_{\rm{ZN}}$ obtained in TGMC, as described in the next paragraph.

The Glauber Monte Carlo is tuned to describe the experimental distributions of
ZN energy. 
For each 1\% centrality interval, the mean number of spectator
neutrons ($N_s$) is obtained in the Glauber Monte Carlo. Folding the ZN response
yields the simulated values of mean energy as a function of centrality.
The experimentally measured mean energy
in the ZN is also determined for each 1\% centrality interval. The
ratio of the measured value of mean energy to the simulated value of
mean energy gives the fractional loss (\textit f) of neutrons due to spectator
fragments that veer away due to the magnetic field. The value of \textit{f} for
the 0-5\% centrality interval is 0.19. For all other centralities it
varies from 0.40 for 5-10\% to 0.55 for 30-35\% centrality interval. 
\begin{figure}[ht!]
\centering
\includegraphics[trim=0.0cm 0.2cm 0.0cm 1.0cm,clip, width=1.0\textwidth]{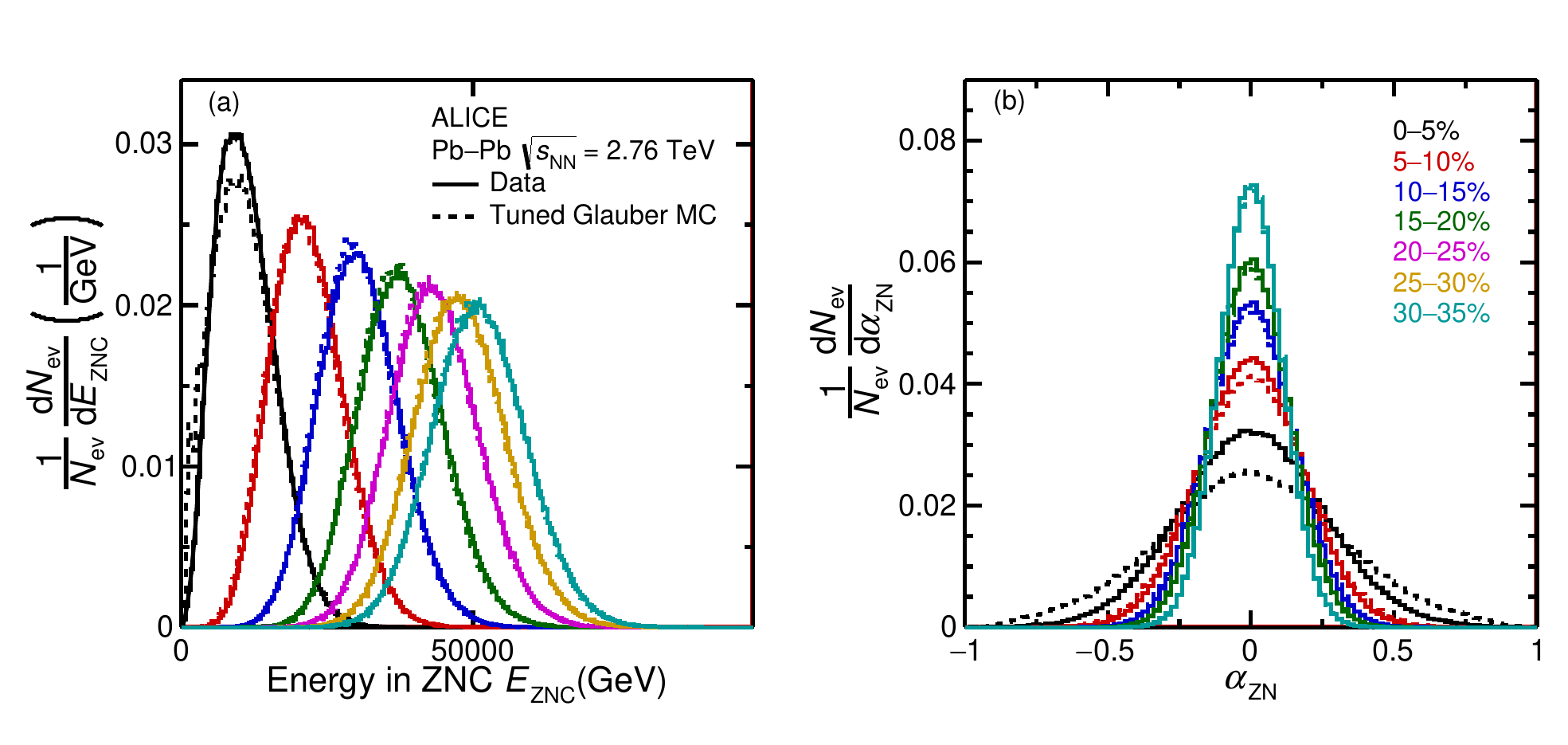}
\caption{(a) Distribution of energy in ZNC in each 5\% centrality
 interval for events simulated using TGMC
 and for the experimental data. (b) Distribution of the asymmetry
 parameter $\alpha_{\rm{ZN}}$ in the simulated events and in experimental data
 for different
 centralities. For clarity, only 5 distributions are shown. The distributions
 corresponding to 20-25\% and 25-30\% lie between those of 15-20\% and 30-35\%.}
\label{fig:ZNDist}
\end{figure} 
A fluctuation proportional to the number of remaining neutrons ($N_s\times (1-f)$)
is incorporated to reproduce the experimental distribution of the energy
deposited in the ZN shown in Fig.~\ref{fig:ZNDist} (a). The peak and
the RMS of the energy distributions
match well. The fractional difference in the position of the peak 
varies between 3.7\% for the 0-5\% centrality interval and 0.1\% for the
30-35\% centrality interval. The fractional difference in
RMS for the most central class is 8.6\% and is in the range 
1.0--2.0\% for all other centrality intervals.
The
distributions of the asymmetry parameter for the TGMC events and the
measured data for each centrality interval are shown in 
Fig.~\ref{fig:ZNDist} (b). The
\begin{figure}[h!]
\begin{center}
\includegraphics[trim= 0.0cm 0.5cm 1.0cm 1.0cm,clip, width=0.5\textwidth]{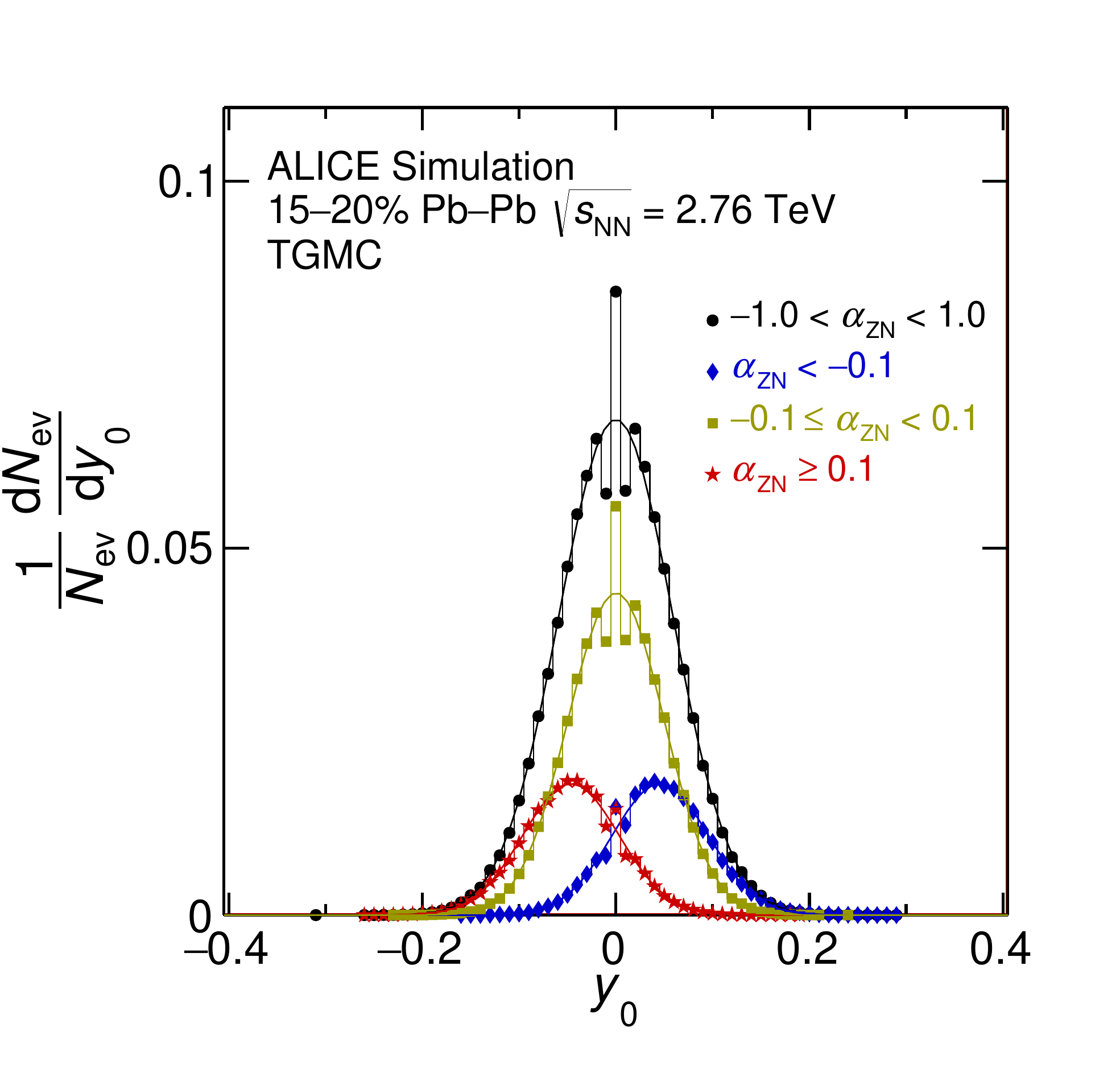}
\caption{The distribution of rapidity-shifts for the events from
  the three different regions of measured asymmetry 
  shown in Fig.~\ref{fig:AlphaDisbnData}. Determination of 
$y_{\rm 0}$ uses the difference in number of nucleons. For small
    values of this difference,  the changes in values near $y_0 = 0$
    are discrete, and are
  smeared into a continuous distribution as $y_{\rm 0}$ increases.}

\label{fig:rapshiftdisbn}
\end{center}
\end{figure} 
\begin{figure}[h!]
\begin{center}
\includegraphics[trim= 0.0cm 0.0cm 0.0cm 0.0cm,clip, width=0.9\textwidth]{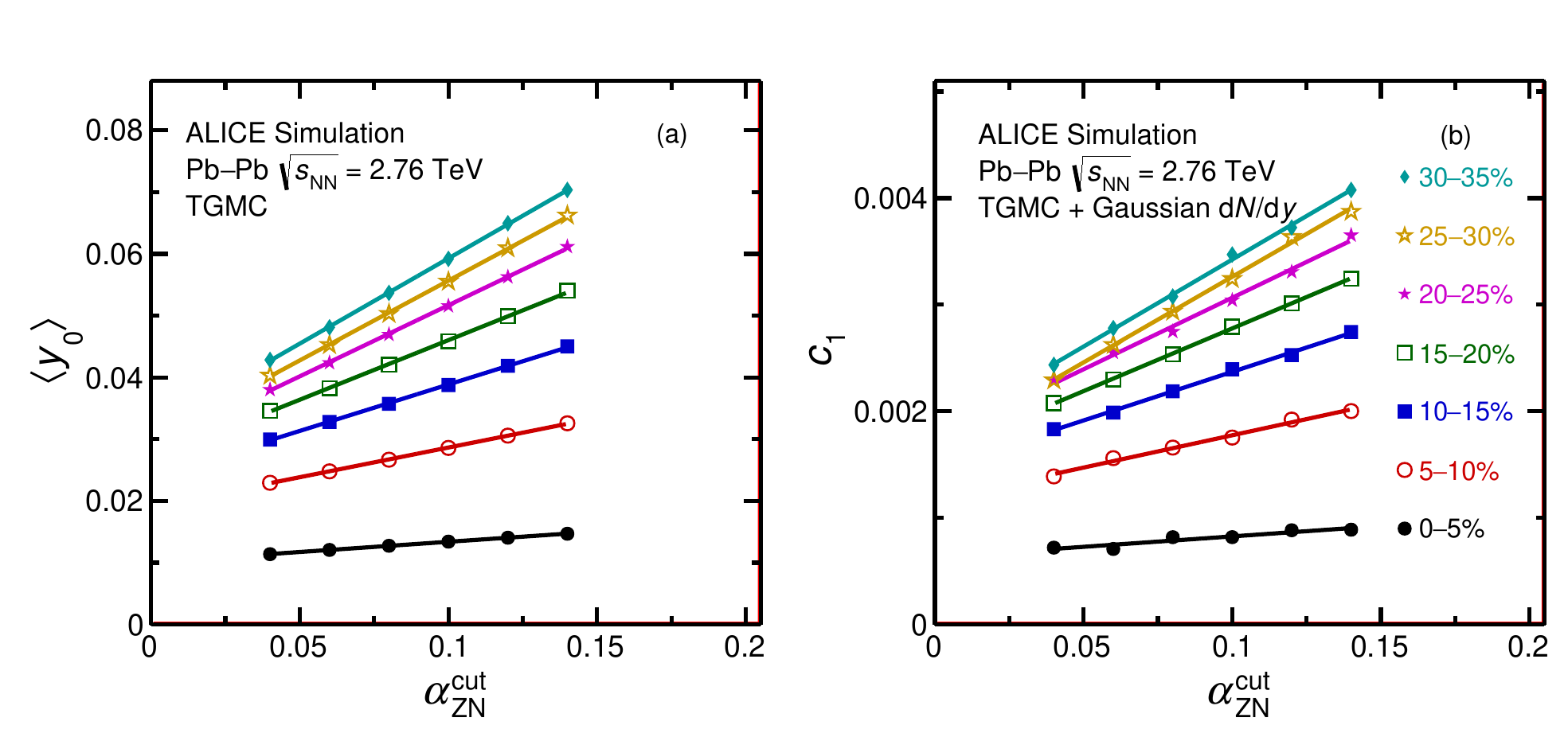}
\caption{(a) The estimated mean value of rapidity-shift $\langle
  y_{\rm 0} \rangle$ for the asymmetric
  region characterised by
  different values of $\alpha_{\rm{ZN}}^{\rm{cut}}$ for each centrality interval. (b)
  The coefficient $c_{\rm 1}$ characterising the change in the
  pseudorapidity distributions for different
  values of $\alpha_{\rm{ZN}}^{\rm{cut}}$, for each centrality interval. These results
  are obtained using TGMC and simulated pseudorapidity
  distributions, as described in the text}.
\label{fig:Meany0vsAsym}
\end{center}
\end{figure} 
TGMC contains information of $y_{\rm 0}$ and $\alpha_{\rm{ZN}}$ for each
event. A scatter plot between $y_{\rm 0}$ and 
$\alpha_{\rm{ZN}}$ is shown in  Fig.~\ref{fig:GlauberMCy0}(d)
for the 15--20\% centrality interval. 
This constitutes the response
matrix. For any measured value of $\alpha_{\rm{ZN}}$, the distribution of
$y_{\rm 0}$ can be obtained. Any difference in the experimental and TGMC
distributions of $\alpha_{\rm {ZN}}$ can be accounted for by scaling the
$y_{\rm 0}$ distribution by the ratio of number of events in data to the
number in TGMC as 
\begin{equation}
f (y_{\rm 0}, \alpha_{\rm {ZN}}^{\rm {Data}}) = f (y_{\rm 0},\alpha_{\rm {ZN}}^{\rm {TGMC}}) \frac{N_{\rm{events}}^{\rm{Data}}}{N_{\rm{events}}^{\rm{TGMC}}},
\label{eq:alphazn}
\end{equation}
with Data (TGMC) in the superscript of number of events,
$N_{\rm{events}}$, denoting the experimental data
(TGMC events). 
For each of the three regions of asymmetry
shown in Fig.~\ref{fig:AlphaDisbnData},
corresponding to a chosen value of $\alpha_{\rm {ZN}}^{\rm {cut}} = 0.1$,
the distribution of rapidity-shift $y_{\rm 0}$ obtained using the response matrix
is shown in Fig.~\ref{fig:rapshiftdisbn}.
 It is worth mentioning that
the width of the distribution of $y_{\rm 0}$ for events from
Region 3, corresponding to 
$-\alpha_{\rm {ZN}}^{\rm {cut}}\leq \alpha_{\rm {ZN}} < \alpha_{\rm {ZN}}^{\rm {cut}}$,
is comparable to the widths of the corresponding distributions
from Regions 1 and 2. The effect of difference in the value of the means of
the $y_{\rm 0}$ distributions is investigated in the present
work.

\subsection{Effect of rapidity-shift on pseudorapidity distributions}\label{rapshift}

The effect of a shift in the rapidity distribution by $y_{\rm 0}$ on the measurable
pseudorapidity distribution ($\dNdeta$) is investigated using simulations.
For each event, the rapidity of charged particles is generated
from a Gaussian distribution of a chosen width \textit{$\sigma_y$}~\cite{Abbas:2013bpa}.
The pseudorapidity is obtained by using the Blast-Wave
model fit to the data for the transverse momentum
 distributions and the experimentally measured relative yields of pions, kaons and protons ~\cite{Abelev:2013vea}.  
To simulate the effect of different widths of the parent rapidity
distribution for different centralities, different $\sigma_{\rm y}$
widths are chosen to
reproduce the measured FWHM (Full Width at Half Maximum) of the pseudorapidity distribution~\cite{Adam:2015kda}.
For the most
central (0--5\%) class, a value 3.86 is used for the width of the rapidity
distribution,  and a value 4.00 is used for the width of the least
central class employed in this analysis (30--35\%). 

The distribution of
rapidity-shift $y_{\rm 0}$, similar to the one shown in Fig.~\ref{fig:rapshiftdisbn}, is
obtained  
for each centrality interval and each $\alpha_{\rm{ZN}}^{\rm{cut}}$ using TGMC. Figure~\ref{fig:Meany0vsAsym} (a) shows the $\langle y_{\rm 0} \rangle$
as a function of $\alpha_{\rm{ZN}}^{\rm{cut}}$ for different
centralities. One observes a linear relation between the two
quantities, showing that an asymmetry in the ZN measurement, 
arising from the unequal number of participating nucleons, is related
to the mean rapidity-shift $\langle y_{\rm 0} \rangle$. 
The rapidity distribution of the particles produced in each event is generated assuming a
Gaussian form centred about a $y_{\rm 0}$, which is generated randomly from the
$y_{\rm 0}$ distribution. 
Events with a rapidity
 distribution shifted by \textit{$y_0$ }$ \neq 0$ yield an asymmetric
 pseudorapidity distribution.
A third order polynomial function in \textit{$\eta$} is fitted to the
ratio of the simulated $\dNdeta$ for the asymmetric region to the
simulated $\dNdeta$ for the symmetric region.
The values of the coefficients in the
 expansion depend upon the rapidity-shift $y_{\rm 0}$ and the
 parameters characterising the distribution ~\cite{Raniwala:2016ugm}.

The simulations described above were repeated for different values of
$\alpha_{\rm{ZN}}^{\rm{cut}}$ to
obtain the pseudorapidity distributions for symmetric and asymmetric
regions. Fitting third order polynomial functions to the ratios of the
simulated pseudorapidity distributions
determines the dependence of ${c_{\rm 1}}$ on  $\alpha_{\rm{ZN}}^{\rm{cut}}$.
Figure~\ref{fig:Meany0vsAsym}(b) shows that ${c_{\rm 1}}$ has a linear dependence on
$\alpha_{\rm{ZN}}^{\rm{cut}}$ for each centrality interval. The
difference in the slopes for
different centralities is
due to differences in the distributions of $y_{\rm 0}$ and to
differences in the widths
of the rapidity distributions.

It is important to note that the parameter ${c_{\rm 1}}$, characterising the asymmetry
in the pseudorapidty distribution, shows a linear dependence
on the parameter $\alpha_{\rm{ZN}}^{\rm{cut}}$ in the 
event sample generated using TGMC and simulations for a Gaussian ${\mathrm{d}N/\mathrm{d}y}$, akin to the dependence of the estimated value of rapidity-shift
$y_{\rm 0}$ for the same sample of events.

\section{Results}
\label{sec:Result}

\begin{figure}[h!]
\begin{center}
\includegraphics[width=0.9\textwidth]{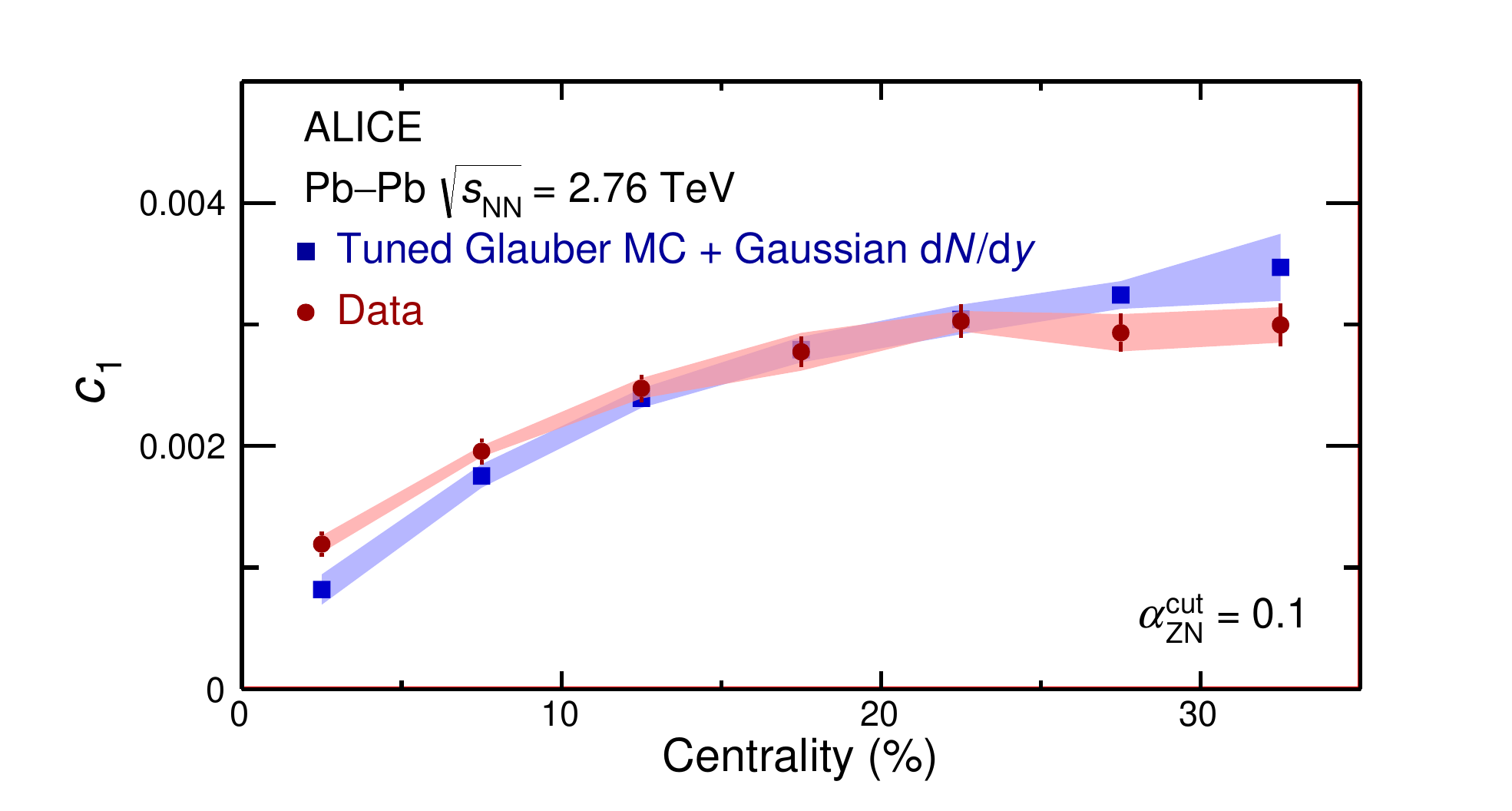}
\caption{The mean values of the coefficient $c_{\rm 1}$ are shown as filled (red)
  circles for different centralities. These correspond to the ratio of
  $\dNdeta$ distributions of populations of events demarcated by
  $\alpha_{\rm{ZN}}^{\rm{cut}}$ = 0.1. The squares
  show the corresponding values from simulations, and correspond to
  $\alpha_{\rm{ZN}}^{\rm{cut}}$ = 0.1 in Fig.~\ref{fig:Meany0vsAsym}, for
  different centralities. The systematic uncertainties are shown as bands.}
\label{fig:Meanc1centrality}
\end{center}
\end{figure}

The longitudinal asymmetry in a heavy-ion collision has been
  estimated from the difference in the energy of the spectator
  neutrons on both sides of the collision vertex.  The effect of the
  longitudinal asymmetry is observed in the ratio of $\dNdeta$
distributions corresponding to different asymmetries. The linear term
in a polynomial fit
to the distribution of the ratio is dominant, and is 
characterised by its coefficient $c_{\rm1}$. 
The centrality dependence of the coefficient $c_{\rm 1}$ for
$\alpha_{\rm {ZN}}^{\rm {cut}} = 0.1$ is shown in
Fig.~\ref{fig:Meanc1centrality}.
\begin{figure}[ht!]
\begin{center}
\includegraphics[trim= 0.0cm 0.0cm 0.2cm 0.0cm,clip, width=0.6\textwidth]{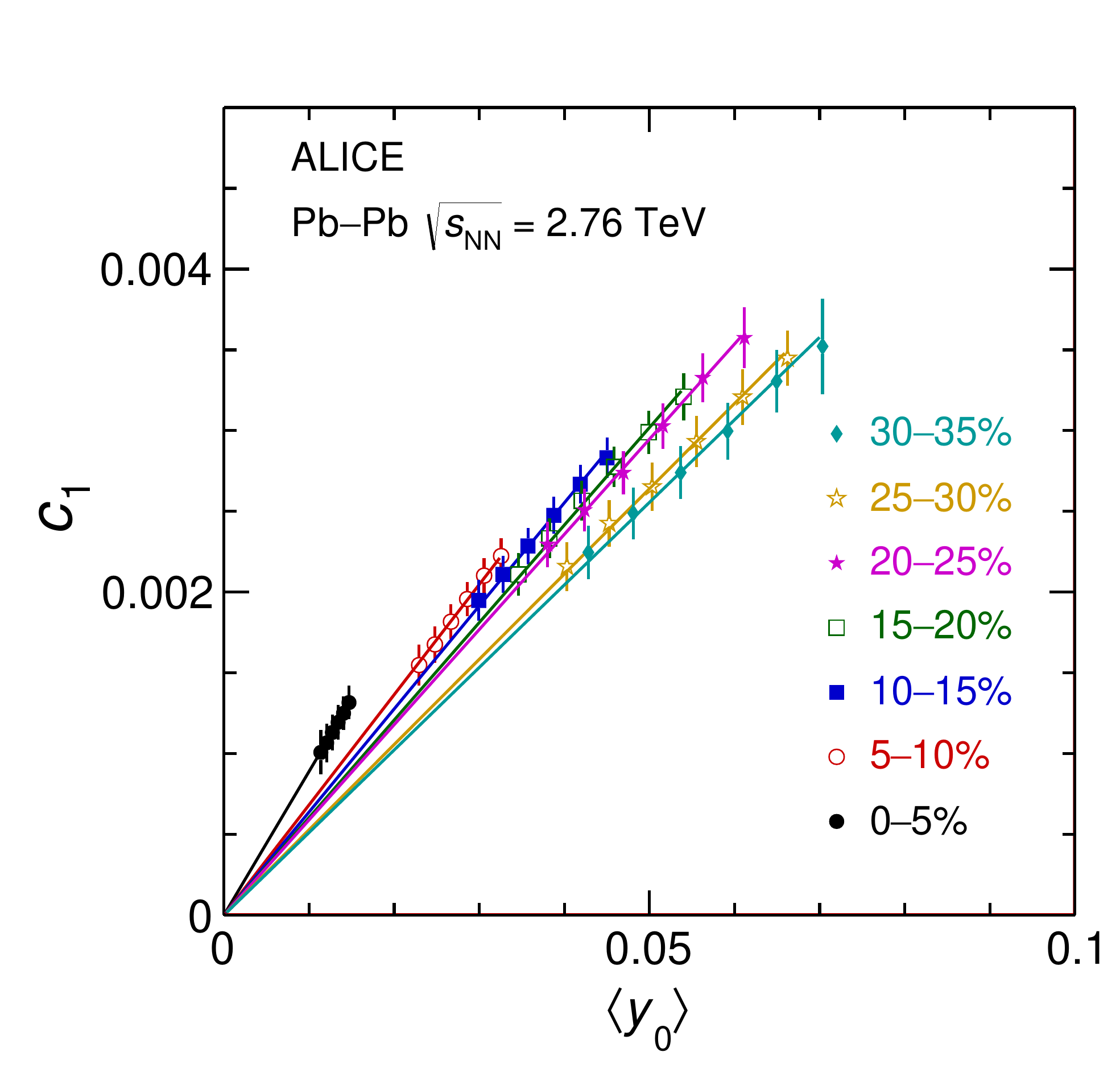}
\caption{For each set of events characterised by $\alpha_{\rm{ZN}}^{\rm{cut}}$,
  the measured values of coefficient $c_{\rm 1}$ as a function of
  estimated values of mean rapidity-shift obtained using TGMC as
  described in the text. The results are shown for
  different centralities. The uncertainties for \textit{$\langle y_0
    \rangle$} shown are statistical and within its symbol size. The lines are
  linear fits passing through the origin.}
\label{fig:c1vsy0}
\end{center}
\end{figure} 
It is worth emphasising that the values of $c_{\rm 1}$ and hence its centrality
dependence are affected by
(i) the distribution of
rapidity-shift $y_{\rm 0}$ for each centrality interval, (ii) the chosen value of
$\alpha_{\rm{ZN}}^{\rm{cut}}$, as seen in Fig.~\ref{fig:Meany0vsAsym}
and (iii) the
shape or the width of the parent rapidity distribution for each
centrality.
Figure~\ref{fig:Meanc1centrality} also shows the results obtained using
simulations as described in Sec.~\ref{rapshift} for
$\alpha_{\rm {ZN}}^{\rm {cut}} = 0.1$. The systematic uncertainty on the 
simulated event sample is estimated by (i) varying the resolution of ZNs from 20\% to 30\%,
(ii) assuming all charged particles are pions and
(iii) varying the width of the parent rapidity distribution within the
  range corresponding to the uncertainties on FWHM quoted in Ref.~\cite{Adam:2015kda}.
The simulated events show a good agreement with the experimental data 
providing credence
to the assumptions of the simulation, in particular that the asymmetry in the
distributions arises from the shift of rapidity of the participant
zone. 

There are two quantities from independent measurements for each selection
of asymmetric events. These are (i) $c_{\rm 1}$, the parameter characterising the
effect of asymmetry in the $\dNdeta$ distributions and shown in
Fig.~\ref{fig:c1alphazdc} and (ii) the mean rapidity-shift $\langle
y_{\rm 0} \rangle$ obtained from the measured asymmetry, filtered
through the corresponding response matrix (Fig.~\ref{fig:GlauberMCy0} (d)), and 
shown in Fig.~\ref{fig:Meany0vsAsym} (a). 
The relation between $c_{\rm
  1}$ and $\langle y_{\rm 0} \rangle$ is shown in  Fig.~\ref{fig:c1vsy0}.
The parameter $c_{\rm 1}$ shows a linear dependence on
\textit{$\langle y_{\rm 0} \rangle$} for each
centrality. 
The difference in the slopes indicates the sensitivity of the
longitudinal asymmetry to the details of the rapidity distribution. 
For a Gaussian rapidity distribution the corresponding parameter $c_{\rm 1}$ 
would be related to the rapidity-shift as $c_{\rm 1} = \frac {y_{\rm 0}}{\sigma_{\rm
    {y^2}}}$ ~\cite{Raniwala:2016ugm}, implying that the slope is
inversely proportional to the square of the width of the
distribution. The observation of an increase in the slope with an increase
in the centrality in the
present data indicates a decrease in the width of the pseudorapidity
distribution with increasing centrality. Such a decrease in the width of
the pseudorapidity distribution with increasing centrality has been
observed independently by fitting the pseudorapidity distributions in
a broad range of pseudorapidity~\cite{Adam:2015kda}.

\section{Conclusions}

The present analysis demonstrates the existence of a longitudinal
asymmetry in the collision of identical nuclei due to fluctuations in
the number of participants from each colliding nucleus. This asymmetry
has been measured in the ZNs in the ALICE experiment (Fig.~\ref{fig:AlphaDisbnData}), and
affects the pseudorapidity distributions, as demonstrated by taking the ratio
of distribution of events from the asymmetric region to the
corresponding one from the symmetric region
(Fig.~\ref{fig:RatiodNdetaV0A}). The effect
can be
characterised by the coefficient of the 
linear term in the polynomial expansion of the ratio. 
The coefficients show a linear
dependence on
$\alpha_{\rm{ZN}}^{\rm{cut}}$, a
parameter to classify the events into symmetric and asymmetric regions
(Fig.~\ref{fig:c1alphazdc}).  Different values of 
$\alpha_{\rm{ZN}}^{\rm{cut}}$ correspond to different values of the
mean rapidity shift $\langle y_{\rm 0} \rangle$ (Fig.~\ref{fig:Meany0vsAsym} (a)).
The parameter describing the change in the pseudorapidity
  distributions ($c_{\rm 1}$) has a simple explanation in the rapidity-shift
  $ \langle y_{\rm 0} \rangle $ of the participant zone (Fig.~\ref{fig:c1vsy0}).
The analysis confirms that the longitudinal distributions are affected
by the rapidity-shift of the participant zone with respect to the
nucleon-nucleon CM frame. 
The results provide support to the relevance of  number of nucleons
affecting the production of charged particles, even at such high
energies.

The longitudinal asymmetry is a good variable to classify the events and provides
information on the initial state of each event. 
A systematic study of the effects of longitudinal asymmetry on
different observables, e.g. the odd harmonics of anisotropic flow, the
forward-backward correlations, the source sizes, in heavy-ion collisions may reveal other characteristics of
the initial state and of particle production phenomena. 

\newpage



\section{Acknowledgements}
\input{fa_2017-10-16.tex}  

\bibliographystyle{utphys}
\bibliography{biblio}{}
\section{The ALICE Collaboration}
\label{app:collab}
\input{Alice_Authorlist_2017-Aug-22.tex}  
\end{document}

%% file: fa_2017-10-16.tex

The ALICE Collaboration would like to thank all its engineers and technicians for their invaluable contributions to the construction of the experiment and the CERN accelerator teams for the outstanding performance of the LHC complex.
The ALICE Collaboration gratefully acknowledges the resources and support provided by all Grid centres and the Worldwide LHC Computing Grid (WLCG) collaboration.
The ALICE Collaboration acknowledges the following funding agencies for their support in building and running the ALICE detector:
A. I. Alikhanyan National Science Laboratory (Yerevan Physics Institute) Foundation (ANSL), State Committee of Science and World Federation of Scientists (WFS), Armenia;
Austrian Academy of Sciences and Nationalstiftung f\"{u}r Forschung, Technologie und Entwicklung, Austria;
Ministry of Communications and High Technologies, National Nuclear Research Center, Azerbaijan;
Conselho Nacional de Desenvolvimento Cient\'{\i}fico e Tecnol\'{o}gico (CNPq), Universidade Federal do Rio Grande do Sul (UFRGS), Financiadora de Estudos e Projetos (Finep) and Funda\c{c}\~{a}o de Amparo \`{a} Pesquisa do Estado de S\~{a}o Paulo (FAPESP), Brazil;
Ministry of Science \& Technology of China (MSTC), National Natural Science Foundation of China (NSFC) and Ministry of Education of China (MOEC) , China;
Ministry of Science, Education and Sport and Croatian Science Foundation, Croatia;
Ministry of Education, Youth and Sports of the Czech Republic, Czech Republic;
The Danish Council for Independent Research | Natural Sciences, the Carlsberg Foundation and Danish National Research Foundation (DNRF), Denmark;
Helsinki Institute of Physics (HIP), Finland;
Commissariat \`{a} l'Energie Atomique (CEA) and Institut National de Physique Nucl\'{e}aire et de Physique des Particules (IN2P3) and Centre National de la Recherche Scientifique (CNRS), France;
Bundesministerium f\"{u}r Bildung, Wissenschaft, Forschung und Technologie (BMBF) and GSI Helmholtzzentrum f\"{u}r Schwerionenforschung GmbH, Germany;
General Secretariat for Research and Technology, Ministry of Education, Research and Religions, Greece;
National Research, Development and Innovation Office, Hungary;
Department of Atomic Energy Government of India (DAE), Department of Science and Technology, Government of India (DST), University Grants Commission, Government of India (UGC) and Council of Scientific and Industrial Research (CSIR), India;
Indonesian Institute of Science, Indonesia;
Centro Fermi - Museo Storico della Fisica e Centro Studi e Ricerche Enrico Fermi and Istituto Nazionale di Fisica Nucleare (INFN), Italy;
Institute for Innovative Science and Technology , Nagasaki Institute of Applied Science (IIST), Japan Society for the Promotion of Science (JSPS) KAKENHI and Japanese Ministry of Education, Culture, Sports, Science and Technology (MEXT), Japan;
Consejo Nacional de Ciencia (CONACYT) y Tecnolog\'{i}a, through Fondo de Cooperaci\'{o}n Internacional en Ciencia y Tecnolog\'{i}a (FONCICYT) and Direcci\'{o}n General de Asuntos del Personal Academico (DGAPA), Mexico;
Nederlandse Organisatie voor Wetenschappelijk Onderzoek (NWO), Netherlands;
The Research Council of Norway, Norway;
Commission on Science and Technology for Sustainable Development in the South (COMSATS), Pakistan;
Pontificia Universidad Cat\'{o}lica del Per\'{u}, Peru;
Ministry of Science and Higher Education and National Science Centre, Poland;
Korea Institute of Science and Technology Information and National Research Foundation of Korea (NRF), Republic of Korea;
Ministry of Education and Scientific Research, Institute of Atomic Physics and Romanian National Agency for Science, Technology and Innovation, Romania;
Joint Institute for Nuclear Research (JINR), Ministry of Education and Science of the Russian Federation and National Research Centre Kurchatov Institute, Russia;
Ministry of Education, Science, Research and Sport of the Slovak Republic, Slovakia;
National Research Foundation of South Africa, South Africa;
Centro de Aplicaciones Tecnol\'{o}gicas y Desarrollo Nuclear (CEADEN), Cubaenerg\'{\i}a, Cuba, Ministerio de Ciencia e Innovacion and Centro de Investigaciones Energ\'{e}ticas, Medioambientales y Tecnol\'{o}gicas (CIEMAT), Spain;
Swedish Research Council (VR) and Knut \& Alice Wallenberg Foundation (KAW), Sweden;
European Organization for Nuclear Research, Switzerland;
National Science and Technology Development Agency (NSDTA), Suranaree University of Technology (SUT) and Office of the Higher Education Commission under NRU project of Thailand, Thailand;
Turkish Atomic Energy Agency (TAEK), Turkey;
National Academy of  Sciences of Ukraine, Ukraine;
Science and Technology Facilities Council (STFC), United Kingdom;
National Science Foundation of the United States of America (NSF) and United States Department of Energy, Office of Nuclear Physics (DOE NP), United States of America.

%% file: Alice_Authorlist_2017-Aug-22.tex

\begingroup
\small
\begin{flushleft}
S.~Acharya\Irefn{org137}\And 
J.~Adam\Irefn{org96}\And 
D.~Adamov\'{a}\Irefn{org93}\And 
J.~Adolfsson\Irefn{org32}\And 
M.M.~Aggarwal\Irefn{org98}\And 
G.~Aglieri Rinella\Irefn{org33}\And 
M.~Agnello\Irefn{org29}\And 
N.~Agrawal\Irefn{org46}\And 
Z.~Ahammed\Irefn{org137}\And 
N.~Ahmad\Irefn{org15}
% ...
\And
\renewcommand\labelenumi{\textsuperscript{\theenumi}~}

\section*{Affiliation notes}
\renewcommand\theenumi{\roman{enumi}}
\begin{Authlist}
\item \Adef{org*}Deceased
\item \Adef{orgI}Dipartimento DET del Politecnico di Torino, Turin, Italy
\item \Adef{orgII}Georgia State University, Atlanta, Georgia, United States
\item \Adef{orgIII}M.V. Lomonosov Moscow State University, D.V. Skobeltsyn Institute of Nuclear, Physics, Moscow, Russia
\item \Adef{orgIV}Department of Applied Physics, Aligarh Muslim University, Aligarh, India
\item \Adef{orgV}Institute of Theoretical Physics, University of Wroclaw, Poland
\end{Authlist}

\section*{Collaboration Institutes}
\renewcommand\theenumi{\arabic{enumi}~}
\begin{Authlist}
\item \Idef{org1}A.I. Alikhanyan National Science Laboratory (Yerevan Physics Institute) Foundation, Yerevan, Armenia
\item \Idef{org2}Benem\'{e}rita Universidad Aut\'{o}noma de Puebla, Puebla, Mexico
\item \Idef{org3}Bogolyubov Institute for Theoretical Physics, Kiev, Ukraine
\item \Idef{org4}Bose Institute, Department of Physics  and Centre for Astroparticle Physics and Space Science (CAPSS), Kolkata, India
\item \Idef{org5}Budker Institute for Nuclear Physics, Novosibirsk, Russia
\item \Idef{org6}California Polytechnic State University, San Luis Obispo, California, United States
\item \Idef{org7}Central China Normal University, Wuhan, China
\item \Idef{org8}Centre de Calcul de l'IN2P3, Villeurbanne, Lyon, France
\item \Idef{org9}Centro de Aplicaciones Tecnol\'{o}gicas y Desarrollo Nuclear (CEADEN), Havana, Cuba
\item \Idef{org10}Centro de Investigaci\'{o}n y de Estudios Avanzados (CINVESTAV), Mexico City and M\'{e}rida, Mexico
\item \Idef{org11}Centro Fermi - Museo Storico della Fisica e Centro Studi e Ricerche ``Enrico Fermi', Rome, Italy
\item \Idef{org12}Chicago State University, Chicago, Illinois, United States
\item \Idef{org13}China Institute of Atomic Energy, Beijing, China
\item \Idef{org14}COMSATS Institute of Information Technology (CIIT), Islamabad, Pakistan
\item \Idef{org15}Department of Physics, Aligarh Muslim University, Aligarh, India
\item \Idef{org16}Department of Physics, Ohio State University, Columbus, Ohio, United States
\item \Idef{org17}Department of Physics, Pusan National University, Pusan, Republic of Korea
\item \Idef{org18}Department of Physics, Sejong University, Seoul, Republic of Korea
\item \Idef{org19}Department of Physics, University of Oslo, Oslo, Norway
\item \Idef{org20}Department of Physics and Technology, University of Bergen, Bergen, Norway
\item \Idef{org21}Dipartimento di Fisica dell'Universit\`{a} 'La Sapienza' and Sezione INFN, Rome, Italy
\item \Idef{org22}Dipartimento di Fisica dell'Universit\`{a} and Sezione INFN, Cagliari, Italy
\item \Idef{org23}Dipartimento di Fisica dell'Universit\`{a} and Sezione INFN, Trieste, Italy
\item \Idef{org24}Dipartimento di Fisica dell'Universit\`{a} and Sezione INFN, Turin, Italy
\item \Idef{org25}Dipartimento di Fisica e Astronomia dell'Universit\`{a} and Sezione INFN, Bologna, Italy
\item \Idef{org26}Dipartimento di Fisica e Astronomia dell'Universit\`{a} and Sezione INFN, Catania, Italy
\item \Idef{org27}Dipartimento di Fisica e Astronomia dell'Universit\`{a} and Sezione INFN, Padova, Italy
\item \Idef{org28}Dipartimento di Fisica `E.R.~Caianiello' dell'Universit\`{a} and Gruppo Collegato INFN, Salerno, Italy
\item \Idef{org29}Dipartimento DISAT del Politecnico and Sezione INFN, Turin, Italy
\item \Idef{org30}Dipartimento di Scienze e Innovazione Tecnologica dell'Universit\`{a} del Piemonte Orientale and INFN Sezione di Torino, Alessandria, Italy
\item \Idef{org31}Dipartimento Interateneo di Fisica `M.~Merlin' and Sezione INFN, Bari, Italy
\item \Idef{org32}Division of Experimental High Energy Physics, University of Lund, Lund, Sweden
\item \Idef{org33}European Organization for Nuclear Research (CERN), Geneva, Switzerland
\item \Idef{org34}Excellence Cluster Universe, Technische Universit\"{a}t M\"{u}nchen, Munich, Germany
\item \Idef{org35}Faculty of Engineering, Bergen University College, Bergen, Norway
\item \Idef{org36}Faculty of Mathematics, Physics and Informatics, Comenius University, Bratislava, Slovakia
\item \Idef{org37}Faculty of Nuclear Sciences and Physical Engineering, Czech Technical University in Prague, Prague, Czech Republic
\item \Idef{org38}Faculty of Science, P.J.~\v{S}af\'{a}rik University, Ko\v{s}ice, Slovakia
\item \Idef{org39}Faculty of Technology, Buskerud and Vestfold University College, Tonsberg, Norway
\item \Idef{org40}Frankfurt Institute for Advanced Studies, Johann Wolfgang Goethe-Universit\"{a}t Frankfurt, Frankfurt, Germany
\item \Idef{org41}Gangneung-Wonju National University, Gangneung, Republic of Korea
\item \Idef{org42}Gauhati University, Department of Physics, Guwahati, India
\item \Idef{org43}Helmholtz-Institut f\"{u}r Strahlen- und Kernphysik, Rheinische Friedrich-Wilhelms-Universit\"{a}t Bonn, Bonn, Germany
\item \Idef{org44}Helsinki Institute of Physics (HIP), Helsinki, Finland
\item \Idef{org45}Hiroshima University, Hiroshima, Japan
\item \Idef{org46}Indian Institute of Technology Bombay (IIT), Mumbai, India
\item \Idef{org47}Indian Institute of Technology Indore, Indore, India
\item \Idef{org48}Indonesian Institute of Sciences, Jakarta, Indonesia
\item \Idef{org49}INFN, Laboratori Nazionali di Frascati, Frascati, Italy
\item \Idef{org50}INFN, Laboratori Nazionali di Legnaro, Legnaro, Italy
\item \Idef{org51}INFN, Sezione di Bari, Bari, Italy
\item \Idef{org52}INFN, Sezione di Bologna, Bologna, Italy
\item \Idef{org53}INFN, Sezione di Cagliari, Cagliari, Italy
\item \Idef{org54}INFN, Sezione di Catania, Catania, Italy
\item \Idef{org55}INFN, Sezione di Padova, Padova, Italy
\item \Idef{org56}INFN, Sezione di Roma, Rome, Italy
\item \Idef{org57}INFN, Sezione di Torino, Turin, Italy
\item \Idef{org58}INFN, Sezione di Trieste, Trieste, Italy
\item \Idef{org59}Inha University, Incheon, Republic of Korea
\item \Idef{org60}Institut de Physique Nucl\'eaire d'Orsay (IPNO), Universit\'e Paris-Sud, CNRS-IN2P3, Orsay, France
\item \Idef{org61}Institute for Nuclear Research, Academy of Sciences, Moscow, Russia
\item \Idef{org62}Institute for Subatomic Physics of Utrecht University, Utrecht, Netherlands
\item \Idef{org63}Institute for Theoretical and Experimental Physics, Moscow, Russia
\item \Idef{org64}Institute of Experimental Physics, Slovak Academy of Sciences, Ko\v{s}ice, Slovakia
\item \Idef{org65}Institute of Physics, Academy of Sciences of the Czech Republic, Prague, Czech Republic
\item \Idef{org66}Institute of Physics, Bhubaneswar, India
\item \Idef{org67}Institute of Space Science (ISS), Bucharest, Romania
\item \Idef{org68}Institut f\"{u}r Informatik, Johann Wolfgang Goethe-Universit\"{a}t Frankfurt, Frankfurt, Germany
\item \Idef{org69}Institut f\"{u}r Kernphysik, Johann Wolfgang Goethe-Universit\"{a}t Frankfurt, Frankfurt, Germany
\item \Idef{org70}Institut f\"{u}r Kernphysik, Westf\"{a}lische Wilhelms-Universit\"{a}t M\"{u}nster, M\"{u}nster, Germany
\item \Idef{org71}Instituto de Ciencias Nucleares, Universidad Nacional Aut\'{o}noma de M\'{e}xico, Mexico City, Mexico
\item \Idef{org72}Instituto de F\'{i}sica, Universidade Federal do Rio Grande do Sul (UFRGS), Porto Alegre, Brazil
\item \Idef{org73}Instituto de F\'{\i}sica, Universidad Nacional Aut\'{o}noma de M\'{e}xico, Mexico City, Mexico
\item \Idef{org74}IRFU, CEA, Universit\'{e} Paris-Saclay, Saclay, France
\item \Idef{org75}iThemba LABS, National Research Foundation, Somerset West, South Africa
\item \Idef{org76}Joint Institute for Nuclear Research (JINR), Dubna, Russia
\item \Idef{org77}Konkuk University, Seoul, Republic of Korea
\item \Idef{org78}Korea Institute of Science and Technology Information, Daejeon, Republic of Korea
\item \Idef{org79}KTO Karatay University, Konya, Turkey
\item \Idef{org80}Laboratoire de Physique Subatomique et de Cosmologie, Universit\'{e} Grenoble-Alpes, CNRS-IN2P3, Grenoble, France
\item \Idef{org81}Lawrence Berkeley National Laboratory, Berkeley, California, United States
\item \Idef{org82}Moscow Engineering Physics Institute, Moscow, Russia
\item \Idef{org83}Nagasaki Institute of Applied Science, Nagasaki, Japan
\item \Idef{org84}National and Kapodistrian University of Athens, Physics Department, Athens, Greece
\item \Idef{org85}National Centre for Nuclear Studies, Warsaw, Poland
\item \Idef{org86}National Institute for Physics and Nuclear Engineering, Bucharest, Romania
\item \Idef{org87}National Institute of Science Education and Research, HBNI, Jatni, India
\item \Idef{org88}National Nuclear Research Center, Baku, Azerbaijan
\item \Idef{org89}National Research Centre Kurchatov Institute, Moscow, Russia
\item \Idef{org90}Niels Bohr Institute, University of Copenhagen, Copenhagen, Denmark
\item \Idef{org91}Nikhef, Nationaal instituut voor subatomaire fysica, Amsterdam, Netherlands
\item \Idef{org92}Nuclear Physics Group, STFC Daresbury Laboratory, Daresbury, United Kingdom
\item \Idef{org93}Nuclear Physics Institute, Academy of Sciences of the Czech Republic, \v{R}e\v{z} u Prahy, Czech Republic
\item \Idef{org94}Oak Ridge National Laboratory, Oak Ridge, Tennessee, United States
\item \Idef{org95}Petersburg Nuclear Physics Institute, Gatchina, Russia
\item \Idef{org96}Physics Department, Creighton University, Omaha, Nebraska, United States
\item \Idef{org97}Physics department, Faculty of science, University of Zagreb, Zagreb, Croatia
\item \Idef{org98}Physics Department, Panjab University, Chandigarh, India
\item \Idef{org99}Physics Department, University of Cape Town, Cape Town, South Africa
\item \Idef{org100}Physics Department, University of Jammu, Jammu, India
\item \Idef{org101}Physics Department, University of Rajasthan, Jaipur, India
\item \Idef{org102}Physikalisches Institut, Eberhard Karls Universit\"{a}t T\"{u}bingen, T\"{u}bingen, Germany
\item \Idef{org103}Physikalisches Institut, Ruprecht-Karls-Universit\"{a}t Heidelberg, Heidelberg, Germany
\item \Idef{org104}Physik Department, Technische Universit\"{a}t M\"{u}nchen, Munich, Germany
\item \Idef{org105}Purdue University, West Lafayette, Indiana, United States
\item \Idef{org106}Research Division and ExtreMe Matter Institute EMMI, GSI Helmholtzzentrum f\"ur Schwerionenforschung GmbH, Darmstadt, Germany
\item \Idef{org107}Rudjer Bo\v{s}kovi\'{c} Institute, Zagreb, Croatia
\item \Idef{org108}Russian Federal Nuclear Center (VNIIEF), Sarov, Russia
\item \Idef{org109}Saha Institute of Nuclear Physics, Kolkata, India
\item \Idef{org110}School of Physics and Astronomy, University of Birmingham, Birmingham, United Kingdom
\item \Idef{org111}Secci\'{o}n F\'{\i}sica, Departamento de Ciencias, Pontificia Universidad Cat\'{o}lica del Per\'{u}, Lima, Peru
\item \Idef{org112}SSC IHEP of NRC Kurchatov institute, Protvino, Russia
\item \Idef{org113}Stefan Meyer Institut f\"{u}r Subatomare Physik (SMI), Vienna, Austria
\item \Idef{org114}SUBATECH, IMT Atlantique, Universit\'{e} de Nantes, CNRS-IN2P3, Nantes, France
\item \Idef{org115}Suranaree University of Technology, Nakhon Ratchasima, Thailand
\item \Idef{org116}Technical University of Ko\v{s}ice, Ko\v{s}ice, Slovakia
\item \Idef{org117}Technical University of Split FESB, Split, Croatia
\item \Idef{org118}The Henryk Niewodniczanski Institute of Nuclear Physics, Polish Academy of Sciences, Cracow, Poland
\item \Idef{org119}The University of Texas at Austin, Physics Department, Austin, Texas, United States
\item \Idef{org120}Universidad Aut\'{o}noma de Sinaloa, Culiac\'{a}n, Mexico
\item \Idef{org121}Universidade de S\~{a}o Paulo (USP), S\~{a}o Paulo, Brazil
\item \Idef{org122}Universidade Estadual de Campinas (UNICAMP), Campinas, Brazil
\item \Idef{org123}Universidade Federal do ABC, Santo Andre, Brazil
\item \Idef{org124}University of Houston, Houston, Texas, United States
\item \Idef{org125}University of Jyv\"{a}skyl\"{a}, Jyv\"{a}skyl\"{a}, Finland
\item \Idef{org126}University of Liverpool, Liverpool, United Kingdom
\item \Idef{org127}University of Tennessee, Knoxville, Tennessee, United States
\item \Idef{org128}University of the Witwatersrand, Johannesburg, South Africa
\item \Idef{org129}University of Tokyo, Tokyo, Japan
\item \Idef{org130}University of Tsukuba, Tsukuba, Japan
\item \Idef{org131}Universit\'{e} Clermont Auvergne, CNRS/IN2P3, LPC, Clermont-Ferrand, France
\item \Idef{org132}Universit\'{e} de Lyon, Universit\'{e} Lyon 1, CNRS/IN2P3, IPN-Lyon, Villeurbanne, Lyon, France
\item \Idef{org133}Universit\'{e} de Strasbourg, CNRS, IPHC UMR 7178, F-67000 Strasbourg, France, Strasbourg, France
\item \Idef{org134}Universit\`{a} degli Studi di Pavia, Pavia, Italy
\item \Idef{org135}Universit\`{a} di Brescia, Brescia, Italy
\item \Idef{org136}V.~Fock Institute for Physics, St. Petersburg State University, St. Petersburg, Russia
\item \Idef{org137}Variable Energy Cyclotron Centre, Kolkata, India
\item \Idef{org138}Warsaw University of Technology, Warsaw, Poland
\item \Idef{org139}Wayne State University, Detroit, Michigan, United States
\item \Idef{org140}Wigner Research Centre for Physics, Hungarian Academy of Sciences, Budapest, Hungary
\item \Idef{org141}Yale University, New Haven, Connecticut, United States
\item \Idef{org142}Yonsei University, Seoul, Republic of Korea
\item \Idef{org143}Zentrum f\"{u}r Technologietransfer und Telekommunikation (ZTT), Fachhochschule Worms, Worms, Germany
\end{Authlist}
\endgroup